\documentclass[aps,pra,reprint,superscriptaddress]{revtex4-2}
\usepackage{amsmath}
\usepackage{amssymb}
\usepackage{graphicx}
\usepackage{dcolumn}
\usepackage{bm}
\usepackage{color}

\begin{document}

\title{Enhancement of spin noise spectroscopy of rubidium atomic ensemble by using of the polarization squeezed light}

\author{Lele BAI $^1$, Lulu ZHANG $^1$, Yongbiao YANG $^1$, Rui CHANG $^1$, Yao QIN $^{1,2}$, Jun HE $^{1,2,3}$, Xin WEN $^{1,4}$, and Junmin WANG $^{1,2,3, *}$ \\ \it {$^1$State Key Laboratory of Quantum Optics and Quantum Optics Devices, and Institute of Opto-Electronics, Shanxi University, Tai Yuan 030006, People’s Republic of China}\\
\it {$^2$Department of Physics, School of Physics and Electronic Engineering, Shanxi University, Tai Yuan 030006, People’s Republic of China}\\
\it {$^3$Collaborative Innovation Center of Extreme Optics, Shanxi University, Tai Yuan 030006, People’s Republic of China}\\
\it {$^4$Department of Physics, Tsinghua University, Beijing 100084, People’s Republic of China}\\
*wwjjmm@sxu.edu.cn}


\date{\today}

\begin{abstract}
We measured the spin noise spectroscopy (SNS) of rubidium atomic ensemble with two different atomic vapor cells (filled with the buffer gases or coated with paraffin film on the inner wall), and demonstrated the enhancement of signal-to-noise ratio (SNR) by using of the polarization squeezed state (PSS) of 795-nm light field with Stokes operator ${{\overset{\Lambda }{\mathop{S}}\,}_{2}}$  squeezed. PSS is prepared by locking the relative phase between the squeezed vacuum state of light obtained by a sub-threshold optical parametric oscillator and the orthogonal polarized local oscillator beam by means of the quantum noise lock. Under the same conditions, PSS can be employed not only to improve SNR, but also to keep the full width at half maximum (FWHM) of SNS unchanged, compared with the case of using polarization coherent state (PCS), and the enhancement of SNR is positively correlated with the squeezing level of PSS. With the increase of probe laser’s power and atomic number density, the SNR and FWHM of SNS will increase correspondingly. With the help of PSS of Stokes operator ${{\overset{\Lambda }{\mathop{S}}\,}_{2}}$, quantum enhancement of both SNR and FWHM of SNS signal has been demonstrated by controlling optical power of the ${{\overset{\Lambda }{\mathop{S}}\,}_{2}}$ polarization squeezed light beam or atomic number density in our experiments.

© 2021 Optica Publishing Group under the terms of the Optica Publishing Group Open Access Publishing Agreement
\end{abstract}

\maketitle
\section{INTRODUCTION}
In any physical system, the inherent random thermal fluctuation and energy change of a certain physical quantity can be reflected when it is measured repeatedly, corresponding statistical variation of average value is called noise. Spin noise (SN), was first proposed by Bloch [1] and verified experimentally by Sleator [2] in the nuclear system, also regarded as the random distribution of atomic electron spin under quasi-thermodynamic equilibrium. Spin noise spectroscopy (SNS), being an optical tool, is used to reveal a number of capabilities specific of nonlinear optics techniques by the change of rotation angle of far-detuned probe laser’s polarization plane, which is manifested as the noise of spin projection on the propagation direction of probe laser [3]. Aleksandrov and Zapasskii achieved the detection of SNS in the alkali gas vapor cell in an undisturbed manner with detuning laser [4]. Crooker et al. observed the SNS of rubidium and potassium atoms under non-resonant conditions [5]. Horn et al. predicted and measured the SNS of rubidium atoms under the condition of resonant optical detection [6]. Ma et al. performed a comparative experiment in rubidium atomic vapor cell with different broading mechanisms [7]. Lucivero et al. demenstrated squeezed-light spin noise spectroscopy[8]. Meanwhile, SNS has unique significance in the field of spin mechanics of semiconductor nanostructures due to the high content of SN in small spin ensemble; see, e.g., reviews in [9,10].

Generally, the physical information obtained in SNS under thermal equilibrium must be related to and constrained by the linear response function of the system, such as signal-to-noise ratio (SNR) of SNS, transverse spin relaxation time of T2, Lande gF factor, and etc, which is determined by the wave dissipation theorem [11]. Calibration of isotope abundance ratio can be carried out according to results of SNS [5]; the magnitude of the strong magnetic field and the nuclear moment can also be calculated accurately by identifying the various peaks’ positions of nonlinear Zeeman results in SNS [12]. The acquisition of the above SNS signals with high quality is the key to obtain the information in these fields. Therefore, the measurement of SNS with high SNR and narrow full width at half maximum (FWHM) has potential significance.

However, the polarization squeezed state as a special quantum resource, can be described by the Stokes operator on the Poincare sphere and characterized by that the noise of a certain Stokes is lower than the standard quantum noise level, and widely used in some precision measurement fields, such as gravitational wave detection and magnetic field measurement [13,14]. The purpose of this work is to apply the 795-nm PSS with high squeezing level to detect the SNS in two types of rubidium atomic ensembles to achieve the measurement of SNS with higher SNR and narrower FWHM compared with the case of using PCS under the same condition, that is, the quantum enhanced measurement of SN signal, which will be helpful to advance the study of the atomic intrinsic properties, and can also be extended to the field of semiconductor materials.
\section{Theoretical analysis}
An atomic ensemble in thermal equilibrium, with a transverse magnetic field of $B$ perpendicular to the propagation direction of the probe light, is called the Voigt configuration. The polarization value of spins is zero after a long period of averaging, due to the random fluctuation of particles’ spin up and down, but it is not zero in bounded time and can be detected. In the case of the length of atomic vapor cell is $L$ and the cross-section area of laser beam is $A$, the effective total number of atoms is $N=N_0AL$, where $N_0$ represents the atomic number density. Its corresponding fluctuation amplitude of magnetization noise in the system is $N^{1/2}$. Induced Faraday rotation signal is ${{\theta }_{F}}=\pi \upsilon L\left( {{n}^{+}}-{{n}^{-}} \right)/c$, where $n_+$ and $n_- $are the refractive indices of atoms to left-circularly($\sigma _+$) and right-circularly($\sigma_-$) polarized light, respectively; $v$ represents the laser frequency and $c$ is the speed of light in vacuum. Faraday rotation fluctuation$\left\langle {{\theta }_{F}}\left( t \right){{\theta }_{F}}\left( 0 \right) \right\rangle $can be used to characterize the magnetization fluctuation$\left\langle {{\text{m}}_{Z}}\left( t \right){{m}_{Z}}\left( 0 \right) \right\rangle $and measured by a polarimeter, which imprints the changes of relative phase between two polarization components onto the probe laser through the magneto-optical interaction with a high number density atomic vapor, shown in figure 1(a). The probe laser exhibits a random oscillation at the Larmor precession frequency of $v_L$, which is determined by the applied external transverse magnetic field (B), i.e ${{\upsilon }_{L}}=B\times \gamma $, where $\gamma$ represents gyromagnetic ratio of atomic ground state. The correlation function of random process has a shape of oscillation attenuation in the time domain with a characteristic time of $T_2$, see figure 1(b). In other words, its geometry is represented by the magnetic resonance spectrum of the spin system. The power spectral density of SN in the frequency domain obtained by Fourier transform is $S\left( \upsilon >0 \right)=\int_{0}^{\infty }{\text{dt}\cos \left( \upsilon t \right)}\left\langle {{\theta }_{F}}\left( t \right){{\theta }_{F}}\left( 0 \right) \right\rangle \propto \frac{\text{1/}{{T}_{2}}}{{{\left( \upsilon -{{\upsilon }_{L}} \right)}^{2}}+1/T_{2}^{2}}$, where $T_2$ is related to laser power (P) and atomic number density (N0), i.e $\Delta \upsilon \text{=}\frac{\text{1}}{\pi {{T}_{\text{2}}}}\text{=}{{\Gamma }_{0}}\text{+}\alpha P\text{+}\beta {{N}_{0}}$, $\gamma_0$ represents the atomic spin relaxation rate in the undisturbed state, $\alpha$ and $\beta$ represent their respective broadening factors. The above discussion is only applicable to atomic vapor cell with homogeneous broadening mechanisms, and corresponding SNS is a Lorentz function centered on the larmor precession frequency in the frequency domain, depicted in figure 1(c). However, SNS with high SNR and narrow FWHM can also be achieved in an atomic vapor cell coated with an anti-relaxation film like octadecyltrichlorosilane [15] on the inner wall due to the reduced destructive effect by wall collisions on atomic spin.
\begin{figure}
\centering
\includegraphics[scale=0.5]{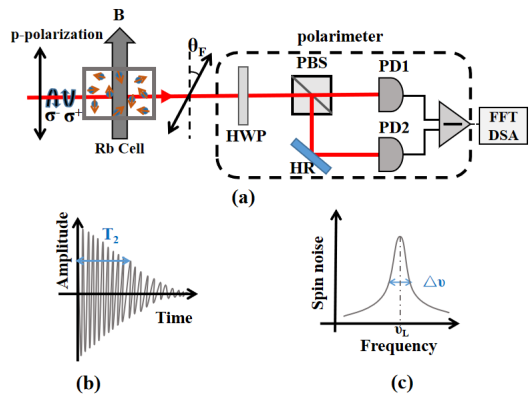}
\caption{\label{Fig_1}Schematic diagram for SNS measurement of Rubidium atomic ensemble. (a) The transverse magnetization of the spin system can be detected with a polarimeter when a p-polarized laser beam passes through a rubidium atomic vapor cell under the driven of transverse magnetic field. In order to avoid the influence of external magnetic field,  atomic vapor cell is placed in the center of a multi-layer permalloy ($\mu$-metal) magnetic shielding tanks. Linearly p-polarization light can be viewed as a superposition state of left-circularly($\sigma_+$) and right-circularly($\sigma_-$) polarized light. The random spin noise fluctuation maps the Faraday rotation angle $\theta_F$ to the polarization direction of the probe laser beam. (b) The stochastic oscillation of probe laser with transverse relaxation time of $\sim$ $T_2$. (c) The spectrum is represented by a Lorentz with the width of $\sim$ $\Delta v$. The transverse magnetic field of B determines the larmor precession frequency of $v_L$, and the FWHM of $\Delta v$ is determined by the dephase time $T_2$. Polarimeter, consisting of a half-wave plate, a polarizing beam splitter and two inversely related photo-detectors, is used to reduce laser classical noise and technology noise and give the final output of differential signal in the measurement of polarization rotation. HWP, half wave plate; PBS, polarization beam splitter; HR, high reflective mirror; PD, photoelectric detector; FFT-DSA, Fast-Fourier-Transform dynamic signal analyzer.}
\end{figure}

Considering the PCS as probe laser, the magnitude of the polarization plane rotation signal output by the polarimeter is characterized by a Fast-Fourier-Transform dynamic signal analyzer (FFT-DSA), and the accompanying output differential voltage can be expressed as:
\begin{equation}
V(t)=G{i_{diff}}(t)
\end{equation}
where $G$ is transimpedance gain of PDs and $i_{diff}$ is the differential current of PD1 and PD2. In the absence of transverse magnetic field, the polarimeter is in equilibrium, where the Faraday rotation angle of $\theta_F$ and the average value of $i_{diff}$ are zero. However, due to the discreteness of photons and its randomness of hitting the detectors, the resulting photo-current shows noise fluctuation:
\begin{equation}
\begin{aligned}
{{(\Delta {{i}_{diff}})}_{PCS}}^{2}&={{(\left\langle {{i}_{diff}}^{2} \right\rangle \text{-}{{\left\langle {{i}_{diff}} \right\rangle }^{2}})}_{PCS}}\\
&={{(\left\langle {{i}_{diff}}^{2} \right\rangle )}_{PCS}}\\
&=2q\left\langle {{i}_{diff}} \right\rangle \Delta \nu
\end{aligned}
\end{equation}
Where $q=1.6 \times 10^{-19}$ C represents the electron charge, $\Delta v$ is the frequency bandwidth of detectors, and the contribution of SNL to the output voltage level is:
\begin{equation}
\begin{aligned}
(\Delta {{V}_{PCS}}{{})^{2}}&=(\Delta {{V}_{SNL}}{{})^{2}}={{\left\langle {{V}^{2}} \right\rangle }_{SNL}}\\
&=2{{G}^{2}}q\left\langle {{i}_{diff}} \right\rangle \Delta \nu =2{{G}^{2}}q\Re P\Delta \nu
\end{aligned}
\end{equation}
Where $\Re$ describes the responsivity of detectors and P refers to the power of laser.

For PSS, the introduction of the squeezing factor ($\xi ^2<1$) makes the contribution of squeezed photon to the background noise as follows:
\begin{equation}
(\Delta {{V}_{PSS}}{{})^{2}}={{\xi }^{\text{2}}}(\Delta {{V}_{PCS}}{{})^{2}}
\end{equation}
and it reduces as the square of squeezing factor decreases, while the size of other signals to be measured will remain unchanged, resulting in the enhanced SNR.
\section{Experimental setup}
In this experiment, the laser source we used is a continuous-wave narrow-linewidth Ti:sapphire laser (M Squared, Model: Solitis), with a tuning wavelength of 700 nm-1000 nm and linewidth of less than 50 kHz, which can satisfy the preparation condition of polarization squeezed light with high squeezing level at the analysis frequency of $\sim$ MHz. The experimental device consists of two parts, as shown in figure 2 (a), one is the preparation of probe laser, which includes PCS and PSS (inside the dashed circle); the other is the establishment of the SNS system (outside the dashed circle). The polarization state of the light field can be described by Stokes parameters on the poincare spheres, see figure 2 (b). In our system, the power noise of different quantum Stokes operator can be measured by the combination of a HWP and a quarter wave plates (QWP), where the Stokes operator ${{\overset{\Lambda }{\mathop{S}}\,}_{2}}$ can be measured by placing a HWP in the balanced homodyne detector (BHD) system when the angle between the axis of HWP and the polarization direction of the incident light field is 21.5°. At this point, if a QWP is placed behind the HWP and the angle set at 45°, then the noise of Stoker operator ${{\overset{\Lambda }{\mathop{S}}\,}_{3}}$ can be measured [16].

Two kinds of laser beams must be focused via a lens before entering the atomic vapor cell to make inherent spin fluctuations more relevant and easier to measure. The probe laser with detuning frequency goes straight through the atomic vapor cell along the z direction, which can prevent the atoms from being excited and disturbed. In the case of a focused probe beam, the shape of the line is independent of the buffer gas pressure [17]. A pair of coaxial Helmholtz coils that produce transverse magnetic field and the atomic vapor cell are placed in a four-layer permalloy ($\mu$-metal) magnetic shield with high magnetic shielding performance to ensure that the atoms are not affected by external uncertainties. The transverse magnetic field is driven by a precise constant current source (Keysight B2961A). Since the SN signal is positively correlated with the atomic number density [18-19], the device of non-magnetic heating and temperature controller are necessary for the atomic vapor cell to ensure uniform distribution of atoms, and the heating frequency is 110 kHz.

The scale of SN signal is determined by the difference in the number of particles spinning up and down in equilibrium, which can result in a Faraday rotation angle of about 100 nrads [20,21]. As the rotation signal of polarization plane is extremely weak and difficult to be identified, the same mode background noise (including system inherent noise and classical noise, etc) should be erased through a polarizer. In our experimental setup, the detection device not only can be used as a BHD system to measure the PSS’s Stokes operator, but also as a polarizer (the same function as in figure 1(a)) to detect the SN signal in the form of differential detection, whose differential signal can be amplified and output to the radio frequency spectrum analyzer (RF-SA) (Agilent, Model: E4405B) or FFT-DSA (Zurich Instruments MFLI-5 MHz), respectively, which can be switched by a single pole dual switch (SPDT). Two vertically polarized beams equally divided by a polarization beam splitter (PBS) are focused on two inversely correlated silicon-based detectors with a common mode rejection ratio of 45 dB (at the analysis frequency of MHz).
\begin{figure}
\centering
\includegraphics[scale=0.5]{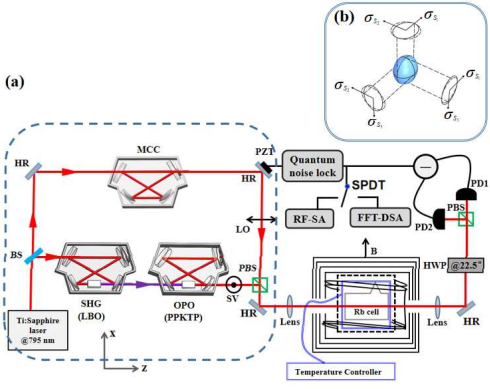}
\caption{\label{Fig_1}Experimental device diagram. The direction of propagation of the probe laser is along z, and the transverse magnetic field along the direction of x. When the output of OPO is blocked, it means that the probe laser of SN meaurement is PCS; while the SV from OPO is combined with LO beam, it represents PSS. The upper right illustration (b) represents the noise ellipsoid of PSS of light prepared in our experiment, which represents Stokes operator of ${{\overset{\Lambda }{\mathop{S}}\,}_{2}}$ is squeezed, Stokes operator of ${{\overset{\Lambda }{\mathop{S}}\,}_{3}}$ is anti-squeezed, and the noise of stokes operator of ${{\overset{\Lambda }{\mathop{S}}\,}_{1}}$ represents SNL. BS, beam splitter; SHG, Second harmonic generation; MCC, Mode clean cavity; HR, high-reflectivity mirror; PZT, Piezoelectric Transducer; LO, the p-polarized local oscillator beam; SV, the s-polarized squeezed vacuum; PBS, polarization beam splitter cube; HWP, half wave plate; RF-SA, the Radio-Frequency spectrum analyzer; FFT-DSA, the Fast-Fourier-Transformation dynamic signal analyzer; SPDT, single-pole double-throw switch.}
\end{figure}
\section{Preparation of 795-nm PSS of light}
The preparation of 795-nm PSS of light mainly consists of four parts: Second harmonic generation (SHG),  optical parametric oscillator (OPO), local oscillator (LO) beam and quantum noise locking (QNL). For simplify the complexity of experimental design, generations of the second harmonic, squeezed vacuum (SV) state of light and LO beam are output from same shape four-mirror ring cavities, which are independent of each other, and have multiple ports for laser injection with great flexibility. For SHG, the nonlinear medium we use is a lithium triborate (LBO) crystal(3 mm$\times$3 mm$\times$13 mm) with the mode of angle matching, the efficiency of frequency doubling is high when the power of fundamental frequency laser is large, even though the nonlinear coefficient is only 0.75 pm/V. Typically, in this experiment, second harmonic of 397.5-nm with the power of 380 mW can be generated under the premise that the power of 795-nm fundamental frequency is 1 W. For OPO, the crystal is periodically poled Potassium titanyl phosphate crystal (PPKTP) (1 mm$\times$2 mm$\times$15 mm) with matching method of phase matching type-0 (e-e+e) and large nonlinear coefficient of $d_{eff}=7\sim 9$ pm/V. However, the transmittance range of wavelength for PPKTP is from $\sim$ 350 nm to $\sim$ 4400 nm, which leads to a particularly severe heat absorption from 397.5nm UV-pumped laser and large loss of the internal cavity, so the squeezing level of 795-nm PSS is severely limited compared with 1064-nm [22].

The OPO cavity is active and the mode of 397.5-nm pumped laser can be realized with quasi phase matching and precise temperature control, while MCC is a passive cavity for high quality spot output to improve the steady high-contrast interference contrast with SV. The length of all above optical cavities are locked by the Pound-drever-Hall technical. The noise of squeezed and anti-squeezed state output by OPO can be expressed as [23,24]:
\begin{equation}
{{R}_{\pm }}\text{=1}\pm \eta {{\varepsilon }^{2}}\zeta \rho \frac{\text{4x}}{{{\left( 1\mp \text{x} \right)}^{2}}+4{{\Omega }^{2}}}
\end{equation}
Where $\eta=94\%$ is the quantum efficiency of the detectors, $\varepsilon=99.7\%$ represents the interference between SV and LO, $\zeta=99\%$ is the laser beam’s transmission efficiency, $\rho=96.6\%$ describes the escape efficiency of OPO cavity, $x=0.63$ is the pump parameters, $\Omega$=0.125 is the tuning parameter in our system and the estimated pumping threshold $P_{th}$ of the OPO is 206 mW. When 397.5-nm pump laser power of OPO is 90 mW, the theoretical relative noise power of SV’s squeezing and anti-squeezing level normalized to the shot noise level are -7.1 dB and +9.4 dB, respectively. The SV state of light is generated by OPO, and corresponding noise power can be obtained by scanning the relative phase between the LO beam and SV [24]. Experimentally, the results of only -5.8$\pm$0.2 dB and +7.3$\pm$0.2 dB under scanning mode are observed at the analysis frequency of 1.63 MHz, as shown in figure 3(a), which are resulted from the decrease of the escape efficiency of OPO cavity, the increase of internal cavity loss and the phase instability, but this is the highest squeezing level prepared experimentally at this wavelength so far.

In SNS measurement, the input laser fields should be continuous PSS of light. Synthesizing the SV (s-polarized) and LO beam (p-polarized) in a certain phase, the PSS of light can be produced after QNL [16, 25], its squeezing level is generally lower than corresponding level measured by scanning mode due to the instability of phase locking. Finally, the relative noise power of stokes operator of ${{\overset{\Lambda }{\mathop{S}}\,}_{2}}$ is -5.3$\pm$0.7 dB, shown in figure 3(b).
\begin{figure}
\centering
\includegraphics[scale=0.4]{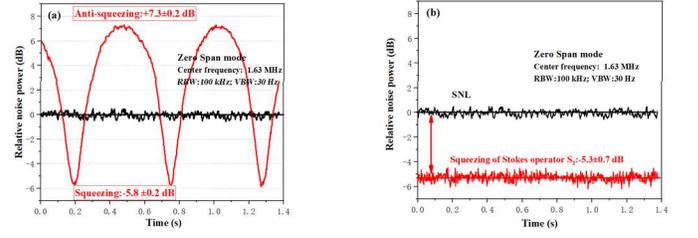}
\caption{\label{Fig_1}Relative noise power spectra normalized to the SNL. (a) The preparation of 795 nm SV state of light. The zero reference line(black) of ordinate represents the SNL. The red curve represents the relative noise power of SV in the scanning mode of LO beam’s phase, and the maximum squeezing and anti-squeezing obtained are -5.8$\pm$0.2 dB and +7.3$\pm$0.2 dB, respectively. (b) The noise of stokes operator ${{\overset{\Lambda }{\mathop{S}}\,}_{2}}$ relative to SNL after quantum noise locking. The noise power spectra are measured with a RF-SA (Aglient E4405B) and averaged for 60 times. The RBW is 100 kHz, VBW is 30 Hz, and the analysis frequency is 1.63 MHz under zero span mode. The Gaussian diameter of the LO beam is 2 mm, optical power is 4 mW. SNL, shot noise level; RBW, resolution bandwidth; VBW, video bandwidth.}
\end{figure}
\section{Quantum enhanced spin noise spectroscopy}
To describe the quality of SNS, it is necessary to normalize its noise power and compare it with the SNL. Rubidium atomic ensemble with natural abundance contains two kinds of isotopic atoms, corresponding SNS measured by FFT-DSA can be fitted by two Lorentz functions, see the schematic diagram of figure 4(a), where the left and right peaks represent the SN signal of $^{85}$Rb and $^{87}$Rb in turn, and two analysis frequencies corresponding to two peaks are their respective Larmor precession frequencies ($v_L$). The red and blue tracks represent the SNS measured by PCS and PSS respectively, and the black dotted line represents the background noise of PCS, also known as the SNL. SNR of SNS can be expressed as:
\begin{equation}
\eta \text{=}\frac{{{S}_{at}}}{{{S}_{\text{ph}}}}
\end{equation}
Where $S_{at}$ is the magnitude of SN signal and $S_{ph}$ is the contribution of laser fields noise proportional to the square of squeezing factor ($\xi$) and optical power ($P$), i.e.,
\begin{equation}
{{S}_{\text{ph}}}\propto {{\xi }^{2}}P
\end{equation}
Among them, for PCS, $\xi=1$, and for PSS, $0<\xi<1$, shown by the red circle and blue ellipse in figure 4(b), respectively. For the results of SNS measurement detected by PCS and PSS with the same power, the background noise of the former is higher than that of the latter, while the signal amplitude of SN remains unchanged, which means that the minimum rotation angle of the polarization plane is more accurate as the noise of PSS’s stokes operator $\hat{S}{}_{2}$ is squeezed, and the SNL can be broken through, so the SNR is improved. S$_{at}$ is proportional to atomic number density and square of probe power, inversely proportional to the effective cross-section area $A$, i.e.,
\begin{equation}
{{S}_{\text{at}}}\propto \frac{{{P}^{2}}{{N}_{0}}L}{A},
\end{equation}

It can be seen from equations (6), (7) and (8) that the increase of laser power and atomic number density are beneficial to improve the SNR of the SN, but they also broaden the FWHM, that is, bring extra power and collision broadening for SNS, see the results of figure 5. Therefore, the trade off of these two conditions is particularly important for SNS measurements with high quality, the introduction of PSS can solve this problem.
\begin{figure}
\centering
\includegraphics[scale=0.4]{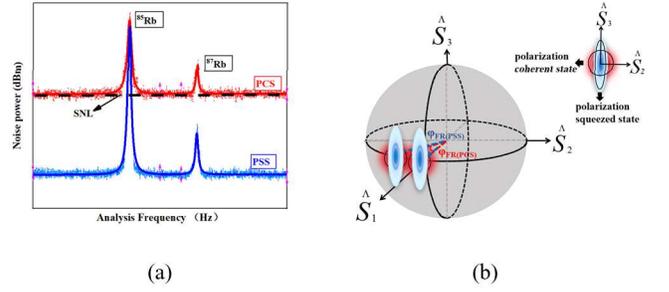}
\caption{\label{Fig_1}Models of two light fields (PCS and PSS) on poincare spheres and their measurement results of SNS. (a) SNS signals are measured by PCS and PSS, respectively. The abscissa and ordinate represent the analysis frequency and noise power respectively. (b) Red sphere represents PCS state, and its polarization will rotate around the equator and has a rotation angle change of $\varphi _{FR(PCS)}$ in the measurement of SNS. The blue ellipsoid represents PSS state, whose Stokes operator$\hat{S}{}_{2}$is squeezed and Stokes operator$\hat{S}{}_{3}$is anti-squeezed. It produces a rotation angle change of $\varphi _{FR(PSS) }$under the same condition.}
\end{figure}

In an atomic vapor cell made in Pyrex material with length of $L=30$ mm and radius of $R=10$ mm, filled with buffer gas at 10 Torr (Ne) and 20 Torr (He), which can slow down the diffusion of atoms to the atomic vapor cell’s inner wall, thus reducing the destruction of the atomic transverse spin relaxation time. When the applied transverse magnetic field is 346.8 $\mu$T and atomic number density is $1.48\times10^{11}/cm^3$ (T=50℃), SNR and FWHM of SN signal are measured at different powers (0.2 mW-6 mW) (only rubidium 85 atoms are discussed here due to it has a large number of atoms in its natural abundance), shown in figure 5(a), where the black solid squares and red solid circles represent the SNR under the measurements of the PCS and the PSS, respectively; the black hollow squares and the red hollow circles represent their respective corresponding FWHM. The SNR is improved at least 3.7 dB (in other words, the signal-to-noise ratio is increased by 2.3 times), and FWHM is unchanged with PSS at the same power compared with PCS. Similarly, as shown in figure 5(b) (the graphics and color represent the same meaning as figure 5(a)), keep the laser power at 3 mW, SNR and FWHM are proportional to the atomic number density (measured from $0.08\times 10^{11}/cm^3$(T=20 ℃) to $5.36\times 10^{11}/cm^3$(T=65 ℃)). The SNR is improved at least 3.9 dB and FWHM is unchanged with PSS at the same atomic number density compared with PCS. According to $\Delta \upsilon \text{=}\frac{\text{1}}{\pi {{T}_{\text{2}}}}\text{=}{{\Gamma }_{0}}\text{+}\alpha P\text{+}\beta {{N}_{0}}$, where $\Delta v$ represents FWHM. The factor of power broadening($\alpha$=3.2 kHz/mW) and collision broadening ($\beta=4.2    
 $\quad$ kHz/10^{11}cm^{-3}$) can be calculated by linearly fitting the corresponding FWHM.
\begin{figure}
\centering
\includegraphics[scale=0.4]{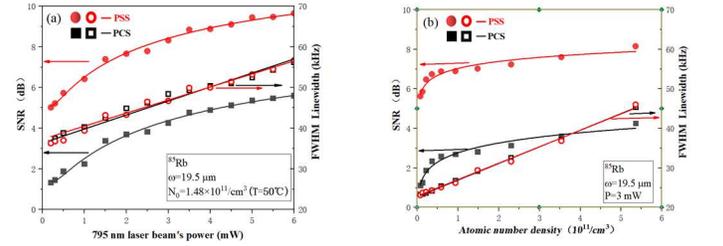}
\caption{\label{Fig_1}The SNR and FWHM of SN are measured by two optical field mechanisms in an atomic vapor cell filled with buffer gas at 10 Torr (Ne) and 20 Torr (He). (a) $N_0=1.48\times 10^{11}/cm^3$, the SNR (5.0-9.7 dB) detected by PSS are higher than the results of PCS (1.3-5.6 dB), and the FWHM (36.4 kHz-56.2 kHz) of PSS and PCS are approximately equal under the same laser power. (b) When the probe laser’s power is 3 mW, the SNR (5.6 dB-8.2 dB) of PSS are also higher than the results measured by PCS (1.1 dB-4.3 dB) under the condition of same atomic number density, and FWHM (23.1 kHz-45.3 kHz) of PSS and PCS are almost same. The solid lines are their respective fitting results.}
\end{figure}

When the atomic number density and the laser power are same for PSS and PCS, typical results are shown in figure 6(a), which show that the the former can improve the SNR of 3.7 dB compared to the later without changing the FWHM. The improved level is slightly lower than the squeezing level of PSS, which is resulted from the destruction of the PSS’s quantum properties after passing through the atomic vapor cell (including the disturbance of high atomic number density, the poor finish of cell’s windows, and the coupling of thermal noise). Based on this, when the power of PSS is reduced to 3 mW, corresponding SNR can still be improved by 1.8 dB, and the low bound of the atomic transverse spin relaxation time ($T_2$) reflected by the FWHM is improved from 5.9 $\mu$s (FWHM $\sim$ 54.3 kHz) to 7.2 $\mu$s (FWHM $\sim$ 44.1 kHz) compared with PCS, as shown in figure 6(b), which benefits from the superiority of quantum enhancement and the reduction of the power broadening. Coincidentally, when the atomic number density measured by PSS goes down by about an order of magnitude, the SNR is improved from $\sim$ 5.4 dB to $\sim$ 6.9 dB compared to PCS, and the value of $T_2$ can also be improved from 5.9 $\mu$s (FWHM $\sim$ 54.3 kHz) to 6.8 $\mu$s (FWHM $\sim$ 46.5 kHz) simultaneously, as shown in figure 6(c), which is due to the fact that the use of PSS not only compensates for the undesirable effect of low atomic number density on the signal of SN, but also reduces the sever collision broadening of SN caused by high atomic number density.
\begin{figure}
\centering
\includegraphics[scale=0.4]{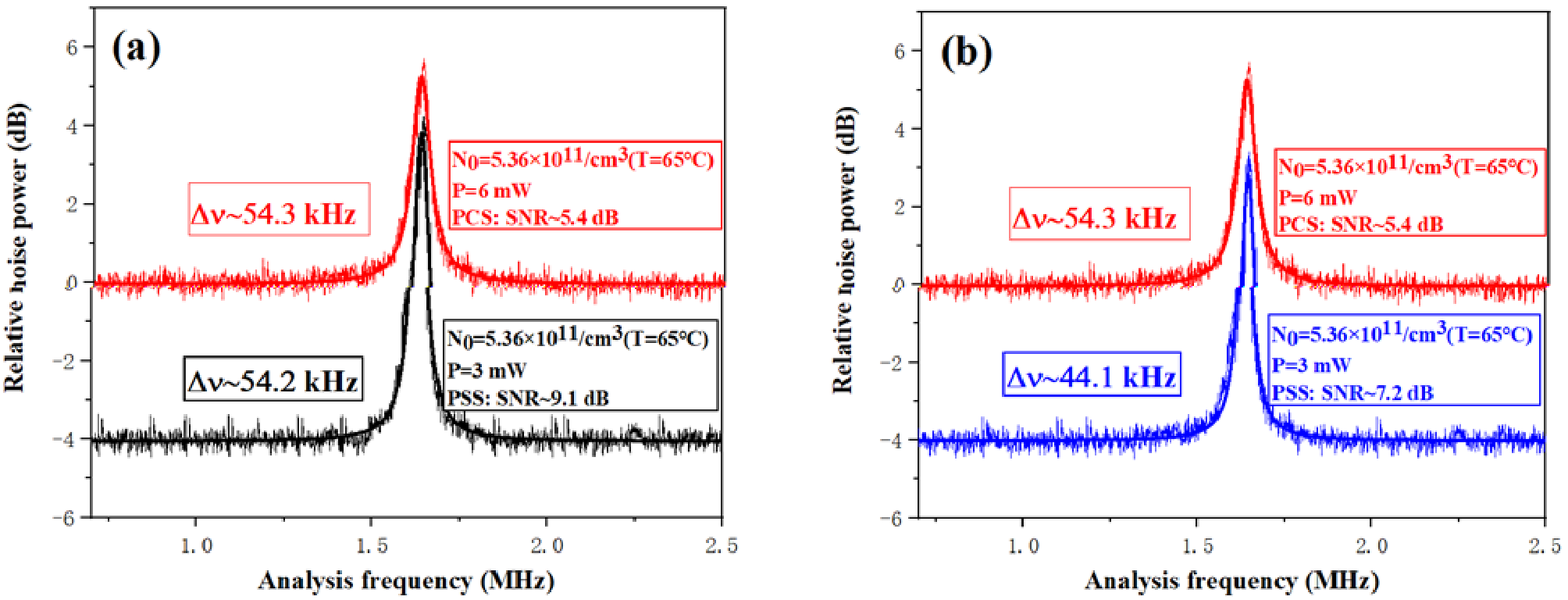}
\includegraphics[scale=0.4]{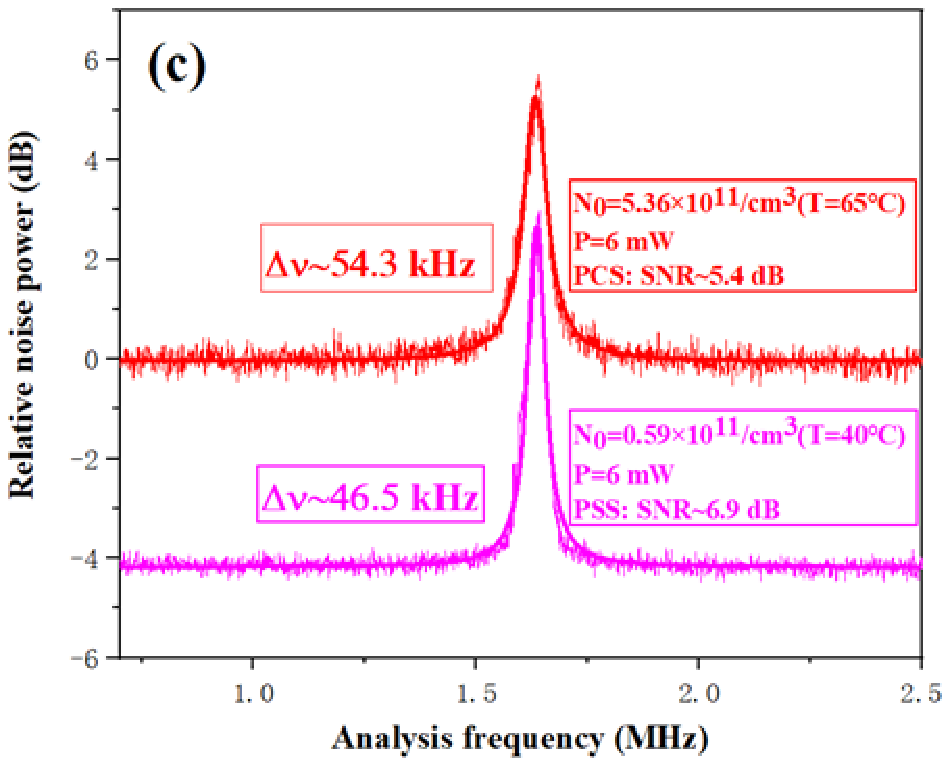}
\caption{\label{Fig_1}Stokes-operator-quantum-state enhanced rubidium atomic SNS in an atomic vapor cell filled with buffer gas at 10 Torr (Ne) and 20 Torr (He). B=346.8 $\mu$ T. All signals are SN of $^{85}$Rb atoms. All red lines represent the SN by using the PCS of light ($N_0$=5.36 $\times 10^{11}/cm^3$, P=6 mW, SNR=5.4 dB, $\Delta$ v=54.3 kHz). (a) The black line is the result of PSS ($N_0$=5.36 $\times 10^{11}cm^3$, P=6 mW, SNR=9.1 dB, $\Delta$ v=54.2 kHz). (b) The blue line is the result of PSS ($N_0$=5.36 $\times 10^{11}/cm^3$, P=3 mW, SNR=7.2 dB, $\Delta$ v=44.1 kHz). (c) The purple line is the result of PSS ($N_0$=0.59 $\times 10^{11}/cm^3$, P=6 mW, SNR=6.9 dB, $\Delta$ v=46.5 kHz). The SNS signals are averaged 200 times.}
\end{figure}

In addition, in order to obtain SNS with narrower FWHM and to prove the universality of this quantum enhancement effect, we applied this method to an another type of atomic vapor cell made in Pyrex material with length of $L=50$ mm and radius of $R=10$ mm, and the inner wall is coated with paraffin film. Finally, we demonstrated that the FWHM was narrowed at the analysis frequency of $\sim$163 kHz, that is, the transverse relaxation time of atomic spins can be well protected. Similarly, the use of PSS of light can still improve the SNR of SNS without changing the FWHM. Typically, when the transverse magnetic field is 34.6 $\mu$ T, the corresponding Larmor precession frequency of $^{85}$Rb is about 163 kHz, and the squeezing level of PSS we use is only -3.7 dB due to the coupling of various noises at this analytical frequency[26, 27]. In order to effectively protect the quantum properties of PSS of light after passing through the atomic vapor cell and to ensure that paraffin film on the inner wall is not damaged [28], the temperature of cell we chose is 45 ℃. Typical results are shown in figure 7. Under the same conditions, when the probe laser power is 1 mW, the atomic number density is $0.94\times 10^{11}/cm^3$, SNR can be improved from 2.4 dB to 5.1 dB compared with the case of using PCS, the FWHM of $\sim$ 6.2 kHz and corresponding low bound of the atomic transverse spin relaxation time of $\sim$ 51.4 µs can be kept unchanged. According to the rules and conclusions obtained in figure 5, the quantum enhancements of both SNR and FWHM can be realized by controlling the laser power of PSS of light and atomic number density. Therefore, we firmly believe that the analysis results can also be extended to other types of atomic vapor cells and semiconductor materials.
\begin{figure}
\centering
\includegraphics[scale=0.4]{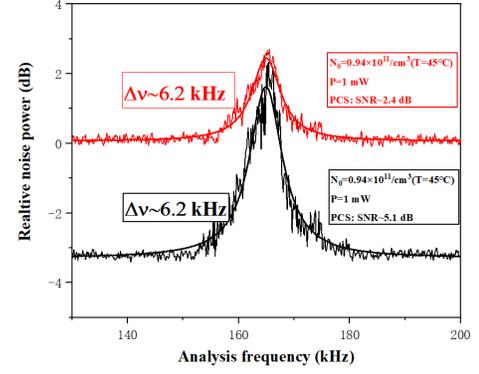}
\caption{\label{Fig_1}Stokes-operator-quantum-state enhanced rubidium atomic SNS in an atomic vapor cell coated with paraffin wax. B=34.6 $\mu$ T. Two signals are SN of $^{85}$Rb atoms. (Red line represents the SNS by using PCS ($N_0$=0.94 $\times 10^{11}/cm^3$, P=1 mW, SNR=2.4 dB, $\Delta$ v=6.2 kHz). The black line is the result of PSS ($N_0$=0.94 $\times 10^{11}/cm^3$, P=1 mW, SNR=5.1 dB, $\Delta$ v=6.2 kHz). The SNS signals are averaged 200 times.}
\end{figure}

\section{Summary and outlook}
The 795-nm PSS with squeezing level of about -5.3$\pm$ 0.7 dB at the analysis frequency of 1.63 MHz was produced by means of OPO. SNS signals in two types of rubidium atomic vapor cells are measured via Faraday rotation with PSS and PCS respectively, and their SNR and FWHM are positively correlated with laser power and atomic number density. Under the same conditions, compared with PCS, PSS has the advantage of breaking SNL to increase SNR without changing the FWHM of SN, which shows the single quantum enhancement properties of PSS. Furthermore, PSS can also improve the both SNR and FWHM simultaneously by appropriately reducing the laser power or atomic number density, which demonstrates the double quantum enhancement properties. These advantages can not only improve the accurate measurement of the intrinsic properties of the atomic ensemble, but also play an important role in the semiconductor field. We also expect that PSS with higher squeezing level has the ability to achieve an order of magnitude of enhanced measurements of weak signals in other precise measurements.

\textbf{Funding.} This work is financially supported by the National Natural Science Foundation of China (Grant Nos. 11974226, 61905133, 11774210, and 61875111), the National Key R\&D Program of China (Grant No. 2017YFA0304502), Shanxi Provincial Graduate Education Innovation Project (Grant No. 2020BY024) and the Shanxi Provincial 1331 Project for Key Subjects Construction.
 
\textbf{Disclosures.} The authors declare no conflicts of interest.

~\\~\\~\\~\\~\\~\\~\\~\\~\\~\\~\\~\\~\\~\\~\\~\\~\\~\\~\\~\\~\\~\\~\\~\\

\end{document}